# Chemical Instability of the Cobalt Oxyhydrate Superconductor under Ambient Conditions

M.L. Foo[1], R.E. Schaak[1], V.L. Miller[1], T. Klimczuk[1,2], N.S. Rogado[1], Yayu Wang[3], G.C. Lau[1], C. Craley[1], H.W. Zandbergen[1,4], N.P. Ong[3], and R.J. Cava[1,5]

[1]Department of Chemistry, Princeton University, Princeton NJ 08544, [2]Faculty of Applied Physics and Mathematics, Gdansk University of Technology, Narutowicza 11/12, 80-952 Gdansk, Poland, [3]Department of Physics, Princeton University, Princeton, NJ 08544, [4]National Centre for HREM, Laboratory of Materials Science, Delft University of Technology, Rotterdamseweg 137, 2628 AL Delft, The Netherlands, [5]Princeton Materials Institute, Princeton University, Princeton NJ 08540.

(April 21, 2003)

## Abstract

The layered sodium cobalt oxyhydrate superconductor $Na_{0.3}CoO_2\cdot 1.4H_2O$ is shown through X-ray diffraction and thermogravimetric studies to be one of a series of hydrated phases of $Na_{0.3}CoO_2$. Further, it is shown that the material is exceptionally sensitive to both temperature and humidity near ambient conditions, easily dehydrating to a non-superconducting lower hydrate. The observation of this stable lower hydrate with c=13.8 Å implies that the superconductivity turns on in this system between $CoO_2$ layer spacings of 6.9 and 9.9 Å at nominally constant chemical doping.

PACs No:74.62.Bf;74.70.-b;74.72Yg

Since the discovery of superconductivity in layered copper oxides in 1986 (1), the search for layered superconductors without copper that might help shed light on the superconducting mechanism of the cuprates has been an important avenue of research. Though other layered oxide superconductors such as $Sr_2RuO_4$ (2) have been reported in the interim, they are clearly fundamentally different from the cuprates both chemically and physically. The recent report of superconductivity near 4K in a layered sodium cobalt oxyhydrate, $Na_{0.35}CoO_2 \cdot 1.4H_2O$, represents the first sign that superconductors related to the cuprates may be found (3). The unusually high Seebeck coefficient of a related compound, $Na_{0.7}CoO_2$ (4), and its attribution to spin entropy (5), suggest that unexpected spin-charge dynamics is indeed at work in this chemical system. Here we report that the oxyhydrate superconductor is one of a series of hydrated phases of $Na_{0.3}CoO_2$. Furthermore it is chemically unstable at body temperature, implying that it cannot be touched during handling without subsequent treatment. It also changes its water content in response to changes in humidity at ambient temperature, implying that it must be handled only under special humidifying conditions. This degree of chemical instability is unprecedented for superconductors, indicating that if physical property measurements are to be performed at all on this important superconductor, extreme attention to ambient conditions will be required.

The crystal structure of the superconducting phase is hexagonal, based on close packed layers of edge-sharing $CoO_2$ octahedra perpendicular to the $c$ axis, separated by intercalant layers consisting of Na and $H_2O$ (see inset to figure 2). There are two layers of $CoO_2$ and intercalant per cell (3). Non-hydrated $Na_xCoO_2$ phases have been of interest for some time (see, for example, ref. 6). Our samples of the superconducting hydrated sodium cobaltate were prepared in a manner similar to that described previously (3) with additional procedures developed to insure high hydration. First, the thermodynamically stable phase $Na_{0.7}CoO_2$ was prepared by the solid-state reaction of $Na_2CO_3$ and $Co_3O_4$ in oxygen at 800°C for 12 hours. A 10% molar excess of $Na_2CO_3$ was added to compensate for loss due to volatilization. $Na_{0.3}CoO_2$ was then prepared by stirring 0.5 g of $Na_{0.7}CoO_2$ in 20 mL of 6M $Br_2$ in acetonitrile at ambient temperature for 1 day to deintercalate sodium. (Similarly, $Na_{0.4}CoO_2$ was prepared by stirring 0.5 g of $Na_{0.7}CoO_2$ in 20 mL of 0.4M $I_2$ in acetonitrile for 1 day.) The product was washed several times with acetonitrile, then with water, and then dried under ambient conditions. The fully hydrated material with the $c=$ 19.6 Å cell reported in reference 3 ($Na_{0.35}CoO_2 \cdot 1.4H_2O$) was prepared by placing 0.5 g of $Na_{0.3}CoO_2$ in an open 20 mL vial, which was then sealed inside a larger screw-cap jar that was loaded with 10 mL of $H_2O$. Evaporation of the water inside the sealed container created a humidified chamber in which samples hydrated over a period of 1-2 days. Exposure of samples to dry air must be minimal after hydration, as described below.

The sodium content of the non-hydrated phases was determined by the inductively coupled plasma (ICP) atomic emission spectroscopy method. The starting material was confirmed to be $Na_{0.7}CoO_2$, the material after the $Br_2$ deintercalation treatment was found to be $Na_{0.3}CoO_2$, and the material after the $I_2$ deintercalation treatment was found to be $Na_{0.4}CoO_2$. The $c$ axis lengths of the non-hydrated $Na_xCoO_2$ phases reveal an expansion of the thickness of the Na intercalant layer with decreased Na content, from 10.82 Å for x = 0.7 to 11.23 Å for x = 0.3. (Because the Na removal results in Co oxidation, the $CoO_2$ layers themselves are expected to shrink.) Attempts to substantially hydrate the $Na_{0.7}CoO_2$ and $Na_{0.4}CoO_2$ powders were not successful under our conditions, though some water was clearly absorbed in both cases.

The change in weight of a highly hydrated, superconducting sample of $Na_{0.3}CoO_2 \cdot yH_2O$ on heating at 0.25 degrees per minute in flowing $O_2$ in a thermogravimetric analyzer (TGA) is presented in figure 1. Powder X-ray diffraction of the initial sample confirmed that it consisted entirely of the hexagonal phase with a $c$ axis parameter of 19.6 Å (Figure 2 shows characteristic regions of the powder X-ray diffraction patterns for this and other phases. For all phases, the hexagonal $a$ axis is approximately 2.83 Å.). On heating to 200°C, the superconducting sample has lost all its water, reverting to $Na_{0.3}CoO_2$. The composition of the $c =$19.6 Å phase from the data in figure 1 can therefore be calculated to be $Na_{0.3}CoO_2 \cdot 1.4H_2O$, similar to that observed in the initial report (3). Figure 1 shows that immediately on heating, $Na_{0.3}CoO_2 \cdot 1.4H_2O$ loses water dramatically. In fact, by body temperature, near 35°C, a clear plateau in the sample weight is observed after a weight loss corresponding to 0.8 $H_2O$ per formula unit. The time dependence of the weight loss of a sample of the fully hydrated phase held at a constant temperature of 35°C in the TGA is shown in the inset to figure 1. Substantial weight loss occurs over a period of several minutes. X-ray diffraction of a sample held at a constant temperature of 35°C for 2 hours showed the complete disappearance of the superconducting $c$ = 19.6 Å phase and the appearance of a partially hydrated phase with $c$ = 13.8 Å. Therefore, superconducting samples lose water and decompose to a non-superconducting lower hydrate at body temperature, indicating that extreme care must be employed to rehydrate samples to observe superconductivity after human contact.

The lower hydrate with the c = 13.8 Å cell formed at the 35°C weight loss plateau (see also figure 2) has not been previously reported. Its formula, based on the data in figure 1, is approximately $Na_{0.3}CoO_2 \cdot 0.6H_2O$. This phase is particularly problematic to maintain in the laboratory, as storage under humid conditions results in hydration to $Na_{0.3}CoO_2 \cdot 1.4H_2O$, and storage in a dry atmosphere results in the formation of $Na_{0.3}CoO_2$. The expansion of the $c$-axis of the cell of $Na_{0.3}CoO_2$ to 13.8 Å in this lower hydrate phase implies an expansion of 1.3 Å per intercalant layer. This is less than the diameter of an oxygen ion (about 2.8 Å), suggesting that in this partial hydrate, the Na ions and $H_2O$ molecules are found in the same plane, unlike the model proposed for the hydrate with the c = 19.6 Å cell (3).

Figure 1 shows the existence of what are likely two more distinct partially hydrated phases that form on heating above 35°C. The first of these is revealed as a relatively broad plateau in weight near 75°C, and the second as a more well defined plateau near 175°C. The water contents of these intermediate hydrates are approximately $Na_{0.3}CoO_2 \cdot 0.3H_2O$ and $Na_{0.3}CoO_2 \cdot 0.1H_2O$, respectively. Powder X-ray diffraction of a fully hydrated sample annealed for several hours at 75°C showed $Na_{0.3}CoO_2 \cdot 0.6H_2O$ with c = 13.8 Å, which we attribute to a rehydration of the phase formed in the 75°C anneal to $Na_{0.3}CoO_2 \cdot 0.6H_2O$ during the X-ray measurement. Similarly, the fully hydrated material annealed at 175 °C for several hours showed a mixture of cells corresponding to $Na_{0.3}CoO_2$ (c=11.2 Å) and $Na_{0.3}CoO_2 \cdot 0.6H_2O$ ($c$=13.8 Å), but with additional broad diffraction peaks corresponding to $c$ values intermediate between 13.8 and 11.2 Å. Again, partial re-hydration during the experiment is implied, though it is as yet impossible to determine this unambiguously. Clear indication of another distinct hydrated phase with intermediate intercalant layer thickness was obtained in several experiments, including one where $Na_{0.3}CoO_2 \cdot 0.6H_2O$ was heated in air at 200 °C for 2 hours. The X-ray pattern for this sample, shown in figure 2, indicates the presence of $Na_{0.3}CoO_2 \cdot 0.6H_2O$ and $Na_{0.3}CoO_2$, and an additional broad (likely indicating an inhomogeneous water content) peak corresponding to a distinct c axis length of 12.7 Å. This phase likely has a water content lower than $0.6H_2O$ per formula unit, based on its observed c axis dimension. Thus the superconducting phase is one of a whole series of hydrates of $Na_{0.3}CoO_2$. Schematic representations of the structures of the most stable of the phases are shown in the inset to figure 2. Isolation of these phases for detailed structural and physical study will be difficult due to the rapid kinetics of hydration and dehydration, but will be of great interest.

The magnetic characterization of the samples on zero field cooling and measuring on heating from 1.9K in a 15Oe applied field is presented in figure 3. The data show that $Na_{0.3}CoO_2$ and $Na_{0.3}CoO_2 \cdot 0.6H_2O$ are not bulk superconductors down to temperatures of 1.9 K. A pure sample of $Na_{0.3}CoO_2 \cdot 1.4H_2O$, with the X-ray diffraction pattern shown in figure 2 marked "19.6 Å superconductor" shows a full diamagnetic screening signal in zero field cooling, and a $T_c$ of 3K. We have not synthesized samples with the 4–5 K $T_c$ reported in reference 3. This may be due to differences in either the degree of hydration or the Na content of the phases. It is of particular interest that synthesizing a hydrated phase with a c-axis parameter of 19.6 Å does not guarantee the creation of a good bulk superconductor. The poorly crystallized (or compositionally inhomogeneous) phase that clearly displays a c= 19.6 Å cell with the XRD pattern shown in figure 2 labeled as "19.6 Å poor superconductor" is not a good bulk superconductor, displaying only the partial screening shown in figure 3. We do not know at the present time whether $H_2O$ deficiency, composition inhomogeneity, or molecular position disorder is the origin of the suppressed superconductivity.

Finally, we show in figure 4 the changes in resistivity of a polycrystalline sample of $Na_{0.3}CoO_2 \cdot xH_2O$ as it is exposed to humid air and then to dry air at ambient temperature. The initial sample is fully dehydrated $Na_{0.3}CoO_2$. On exposure to the humid air (at time t1) the resistivity increases sharply, in a two-step process, as water is absorbed. Our TGA and X-ray diffraction measurements lead us to conclude that the first plateau in increased resistivity corresponds to the formation of $Na_{0.3}CoO_2 \cdot 0.6H_2O$, and the high resistivity plateau corresponds to the formation of $Na_{0.3}CoO_2 \cdot 1.4H_2O$. As the sodium content is constant in the hydration process, and $H_2O$ is formally electrically neutral, then this experiment indicates that a substantial increase in resistivity occurs at constant chemical doping level as the $CoO_2$ layers are increasingly separated by the intercalation of water. The resistivity decreases quickly when the sample is removed from the humidifying chamber (at t2) and exposed to dry air, indicating the rapid disappearance of the fully hydrated phase. Again an intermediate resistivity plateau is observed. The resistivity returns to its initial value on dehydration, but repeated cycles of hydration and dehydration lead to the degradation of the measured resistivity, as is expected if the large volume changes involved eventually create cracks in the sample.

In addition to quantifying the exceptional instability of the superconducting cobalt oxyhydrate superconductor, and suggesting routes for handling the material successfully, our results reveal several basic aspects of the superconductivity. Firstly, at what appears to be constant formal doping, as the separation between $CoO_2$ layers is increased, the room temperature resistivity increases on going from non-superconducting to superconducting phases. Secondly, the observation of the stable intermediate hydrate with c=13.8 Å allows us to conclude that the superconductivity turns on in this system between $CoO_2$ layer spacings of 6.9 and 9.9 Å. It is not known at this time what the optimal spacing between layers and optimal chemical doping level will be. The role of the intercalated $H_2O$ may not be as straightforward as it might first appear. In addition to modifying the spacing of the $CoO_2$ layers, it may cause subtle structural changes in those layers. Further, though formally neutral, it may cause a redistribution of the charge in the $CoO_2$ layer on intercalation. At these high formal oxidation states for Co (3.7+) and in the presence of electropositive $Na^+$, there may be substantial hole character on the oxygens in the $CoO_2$ layer. The additional oxygen that the intercalated water supplies in the bonding sphere of the Na and in the vicinity of the $CoO_2$ layer may cause a redistribution of those doped holes.

**Acknowledgements**

This work was supported by the US National Science Foundation, grants DMR 0244254 and 0213706, and the US Department of Energy, grant DE-FG02-98-ER45706. T. Klimczuk would like to thank The Foundation for Polish Science for support.

**Figure captions**

**Figure 1:** Thermogravimetric analysis of a hydrated sodium cobaltate sample $Na_{0.3}CoO_2 \cdot 1.4H_2O$ with c=19.6 Å – the superconducting cobalt oxyhydrate. Heating rate 0.25°/minute, in flowing bottled $O_2$. Inset: time dependence of the weight loss for $Na_{0.3}CoO_2 \cdot 1.4H_2O$ heated at 0.5°/minute to 35°C in $O_2$ and then held isothermally.

**Figure 2:** Main panel: Powder X-ray diffraction patterns (Cu Kα radiation) of characteristic regions of various $Na_{0.3}CoO_2 \cdot xH_2O$ phases. Inset: schematic representation of the structures of the most stable $Na_{0.3}CoO_2 \cdot xH_2O$ phases.

**Figure 3:** Zero field cooling magnetic characterization of the $Na_{0.3}CoO_2 \cdot xH_2O$ phases (15 Oe DC applied field, PPMS, Quantum Design.).

**Figure 4:** Time dependence of the resistivity (4-probe method, silver paint contacts) of a pressed sample (5GPa, 2.5 hours) of $Na_{0.3}CoO_2 \cdot xH_2O$ when an x=0 sample is exposed to humid air at a time t1, and then to dry air at a time t2.

**Figure 1:**

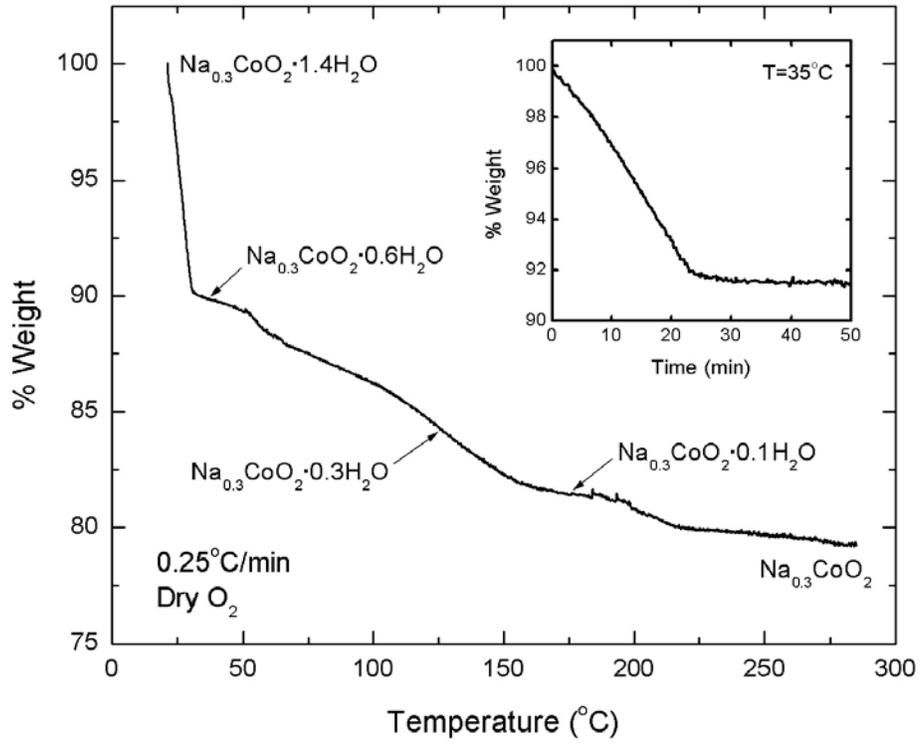

**Figure 2**

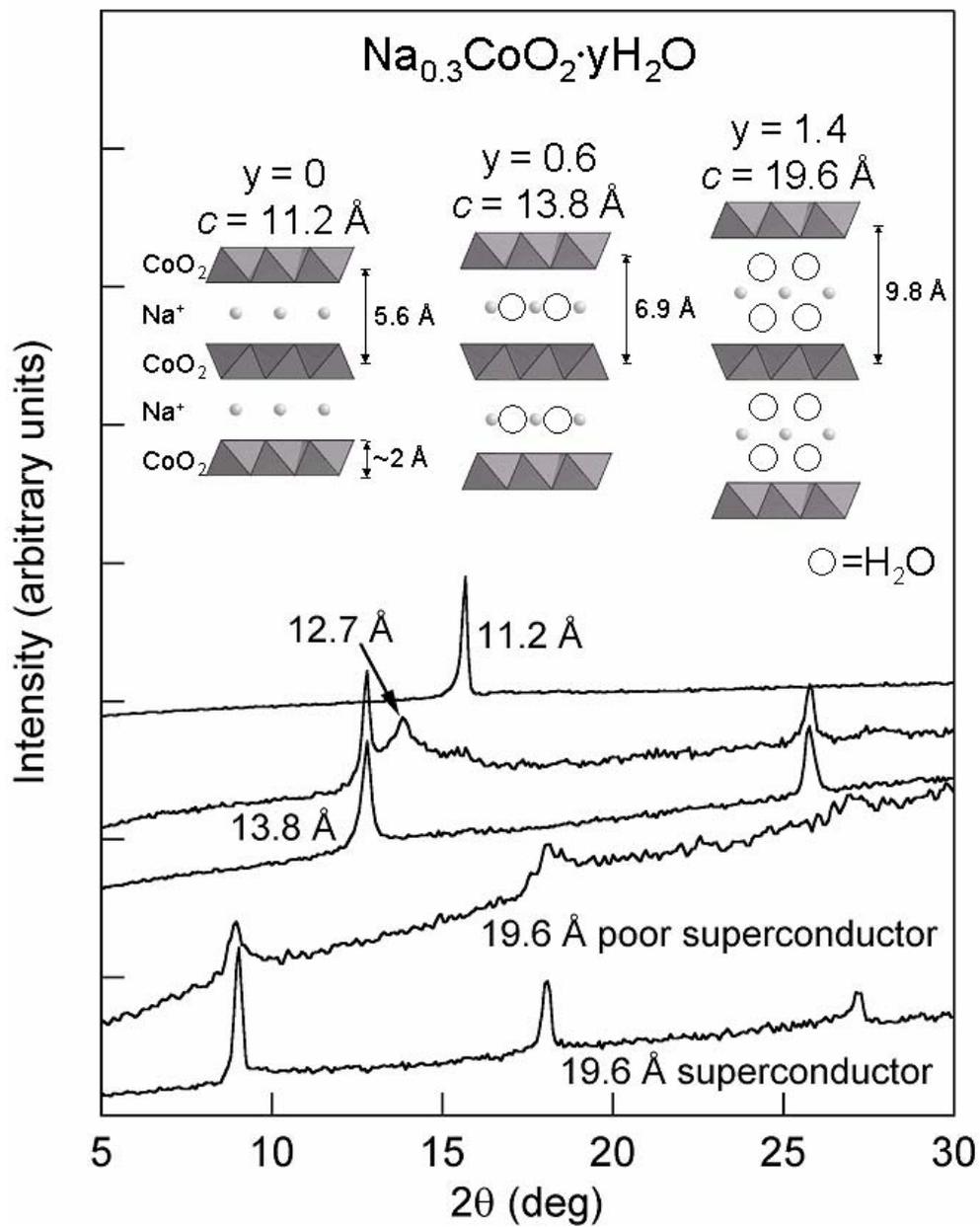

**Figure 3**

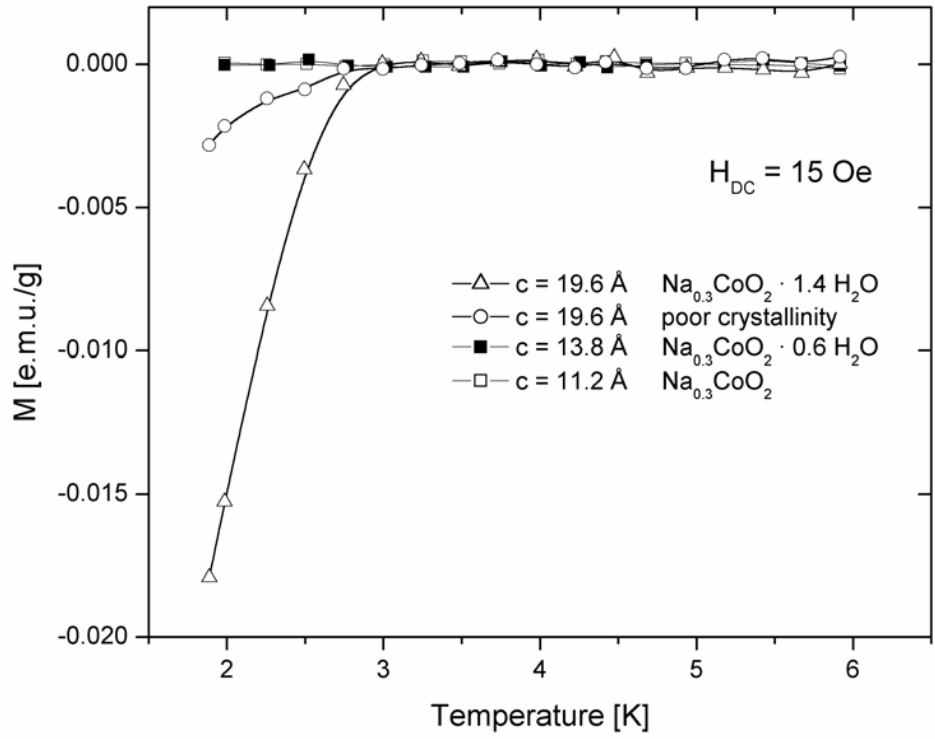

**Figure 4**

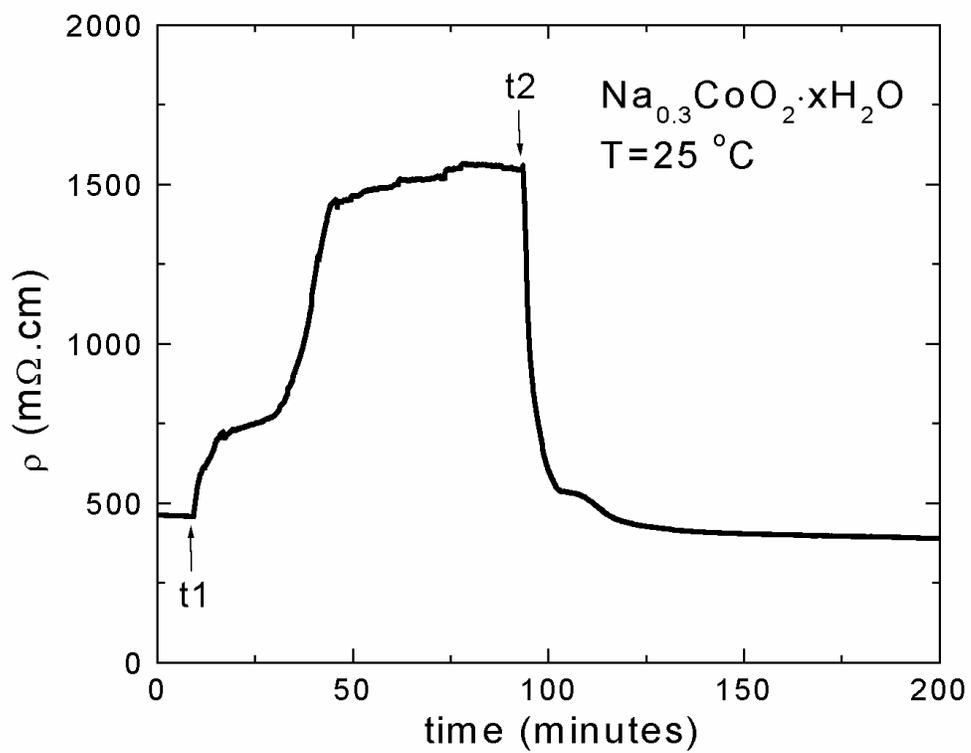